# Misalignment compensation in spatial light modulator based optical filtering techniques

M. Agour[1,2], C. Falldorf[1], C. v. Kopylow[1] and R. B. Bergmann[1]

[1] Bremer Institut für Angewandte Strahltechnik, Klagenfurter Strasse 2, 28359 Bremen, Germany

[2] Physics Department, Aswan Faculty of Science, South Valley University, 81528 Aswan, Egypt

**email:** agour@bias.de

## Summary

A new method for the compensation of misalignment in the spatial light modulator based optical linear filtering techniques is presented. It is based on the correlation of the wave fields generated across the input and the output planes of filtering setups. Experimental results are given to demonstrate the effectiveness of the method.

## Introduction

Spatial light modulator (SLM) based optical filtering techniques, e.g., optical correlators, take the advantage of inherent parallel processing capabilities of optical Fourier transform. The design of these techniques is commonly based on the conventional Vanderlugt filter [1] with an SLM located across the common Fourier plane of the setup. Figure 1 shows a sketch of a 4f-optical filter with a reflective SLM located across the Fourier domain.

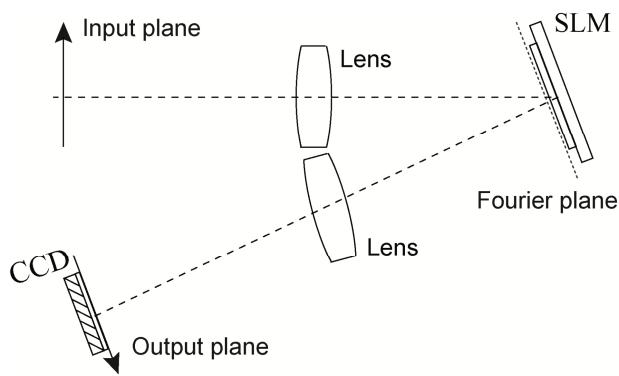

Fig 1. Sketch of a 4f-optiacl filtering setup.

However, these systems suffer from the alignment of the setup's optical axes and the center of the SLM. Let's now describe this setup as being composed of two distinct sections. Each of them consists of an optical axis with a Fourier transforming lens on it. Additionally, these sections are linked by the common Fourier plane where the SLM is located. Recently, we proposed a model describing the effect of misalignment of such setup on the generated wave field across the sensor domain [2]. It is referred to as SLM-based phase retrieval setup. It demonstrates that if misalignment is present, measurement errors will be propagated in the iterative process and thus affect the accuracy and the rate of convergence of the phase retrieval. Based on this model, the setup has been electronically aligned by modifying the transfer function displayed on the SLM. But the alignment method requires a calibration object having specific features e.g. U.S. Air Force resolution target which is characterized by its constant phase [3]. In the following, a new alignment method is proposed.

## Alignment approach

Firstly, using the alignment approach presented in Ref. [3], the setup is aligned i.e. the optical axes of the two sections of the setup hit the center of the SLM. We considered next that the setup is misaligned by a known misalignment in the Fourier plane between section B and the center of the SLM. In this case the wave field generated across the CCD is shifted and modulated with a phase ramp [3]. Since we recently succeed to recover the complex amplitude of a plane wave with the help of the diversity introduced by using a spatial phase diffuser element [4], it means that the ramp term may also be retrieved. Based on the recovered ramp, a new method to align the setup is proposed. Compared to the method proposed in Ref. [3] the new approach does not require any calibration object.

**Experimental results**

In order to verify the presented approach we introduced a known shift of 240 µm at the Fourier plane which corresponds to a shift of 30 SLM's pixels. Consequently, a set of 10 transfer functions centered at (0,30) pixels has been designed and displayed on the SLM. Accordingly, 10 intensity measurements across subsequent planes separated by a distance of $z=2$ mm and in the range from 0 to 18 mm have been captured. The measurements have been subjected into an iterative process [5] and the recovered complex amplitude across the input plane is shown in Fig. 2.

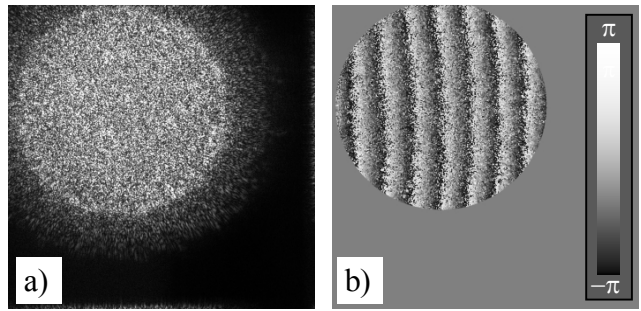

Fig 2. Recovered complex amplitude: amplitude (a) and phase (b) distributions across the input plane of a plane wave for 100 iterations. All distributions: 1024 × 1024 pixel with a pixel pitch of 3.45 µm.

Applying the Fourier transform on the retrieved phase and substituting the setup design parameters $f$=150 mm, $\Delta p$=8 µm and $\lambda$=532 nm, the corresponding shift of 30 pixels in the Fourier domain is calculated. This result equals the shift value which is introduced to the designed transfer function of free space propagation and shows the validity of the new presented method for the alignment of the setup.

**Conclusion**

We have presented a new method for compensation of the misalignment between the optical axes and the center of an SLM of an SLM-based optical filtering setup. It is based on the correlation of the complex amplitudes generated at the input and the output plane of the filter setup. Experimental results demonstrate the effectiveness of our technique.

**References**

[1] J. Goodman, Introduction to Fourier Optics (McGraw-Hill, 1996), 2nd ed.
[2] C. Falldorf, M. Agour, C. von Kopylow, and R. B. Bergmann, *Appl. Opt.*, **49**, 1826-1830, 2010.
[3] M. Agour, C. Falldorf, C. von Kopylow, and R. B. Bergmann, *Appl. Opt.*, **50**, 4779-4787, 2011.
[4] M. Agour, P. Almoro, C. von Kopylow, and C. Falldorf, '*Measurement of thermally induced deformations by means of phase retrieval*', 1[st] EOS Topical Meeting on Micro- and Nano-Optoelectronic Systems, ISBN 978-3-00-033711-6 (2011).

[5] P. Almoro, G. Pedrini, and W. Osten, *Appl. Opt.*, **45**, 8596–8605, 2006.